\documentclass[a4paper,fleqn,usenatbib]{mnras}
\usepackage{newtxtext,newtxmath}
\usepackage[T1]{fontenc}
\usepackage{ae,aecompl}

\usepackage{graphicx}	% Including figure files
\usepackage{color} % For marking changes. To be removed in final version.
\usepackage{breakurl} % breaking url into multiple lines
\usepackage{threeparttable} % For table footnotes
\newcommand{\refmark}{}
\title[Orbital variations in IGR J16318-4848]{Orbital variations in intensity
and spectral properties of the highly
obscured sgHMXB IGR J16318-4848}

\author[N. Iyer et al.]{
  Nirmal Iyer\thanks{E-mail: nirmal.iyer@gmail.com (NI)}\thanks{NI current address :
Albanova University Center, KTH, Stockholm, Sweden, 106 91}, 
Biswajit Paul,  
\\
Raman Research Institute, Bangalore, India 560080 \\
}

\date{Accepted XXX. Received YYY; in original form ZZZ}

\pubyear{2017}

\begin{document}
\label{firstpage}
\pagerange{\pageref{firstpage}--\pageref{lastpage}}
\maketitle

% Abstract of the paper
\begin{abstract}
IGR J16318-4848 is an X-ray binary with the highest known line of sight
absorption column density among all known X-ray binary systems in our
galaxy. In order to investigate the reason behind such a large absorption
column, we looked at the variations in the X-ray intensity and
spectral parameters as a function of the tentatively discovered $\sim$ 80 day
orbit of this source. The orbital period is firmly confirmed in the long term
($\sim$ 12 year) \textit{Swift} \refmark{BAT lightcurve}. 
Two peaks about half an orbit apart, one narrow and small,
and the other broad and large are seen in the orbital intensity profile.
We find that \refmark{while most orbits show enhanced emissions at these two peaks, 
the larger peak in the folded longterm lightcurve is more a} result of randomly occurring large
flares spread over $\sim$ 0.2 orbital phase. As opposed to this, the
smaller peak is seen in every orbit as a regular increase in
intensity. Using archival data spread over different phases of the orbit and
the geometry of the system as known from previously published infrared
observations, we present a possible scenario which explains the orbital
intensity profile, bursting characteristics and large column density
of this X-ray binary.
\end{abstract}

% Select between one and six entries from the list of approved keywords.
% Don't make up new ones.

%%%%%%%%%%%%%%%%%%%%%%%%%%%%%%%%%%%%%%%%%%%%%%%%%%

%%%%%%%%%%%%%%%%% BODY OF PAPER %%%%%%%%%%%%%%%%%%

\section{Introduction}\label{sec:int}
Hard X-ray surveys with the coded mask imagers on \textit{INTEGRAL} and
\textit{Swift} observatories have revealed a class of
highly obscured supergiant high mass X-ray binary systems (sgHMXB) in the
galactic plane. These binaries are characterised by persistently high line of
sight absorption column ($\mathrm{N_H > 10^{23} \, cm^{-2}}$) as inferred
from their X-ray spectrum. Most of these sources are thought to be classical
sgHMXBs undergoing transition to Roche lobe overflow, with low wind
velocities \citep{walt2015}. However, exceptions to such an explanation are
seen in the case of IGR J16320-4751 (with an 8.9 day orbit) and CI Cam (with a 19.4 day
orbit). Such a large orbit makes it unlikely for these systems to be 
Roche lobe transiting systems. IGR J16320-4751 is seen to have significantly larger infrared
reddening than expected from the column estimated via 21-cm HI line surveys, indicating that
this system has a large amount of dust present in the
vicinity of the source \citep{chat2008,walt2015}. CI Cam too is thought to
have the presence of an enshrouding dust shell surrounding the compact object
\citep{bart2013}. The discovery of an 80 day
orbit of IGR J16318-4848 \citep{jain2009}, an X-ray binary with the highest known persistent
absorption column ($\mathrm{N_H \sim 10^{24} \, cm^{-2}}$) \citep{ibar2007} added another
interesting exception to this list.

IGR J16318-4848 was first discovered in 2003 by the IBIS/ISGRI onboard
\textit{INTEGRAL} \citep{cour2003}. It was quickly revealed that this source
had an exceptionally high absorption column and very prominent iron and
nickel fluorescence lines in X-rays \citep{matt2003}. The source was also
seen to display large variations in its X-ray intensity over timescales of
kiloseconds to days \citep{cour2003,krim2010}. Infrared observations post XMM
localization of the source position led to the identification of a possible
sgB[e] counterpart with a high intrinsic absorption and a complex and dense
circumstellar medium surrounding the companion star \citep{fill2004}.
Further spectroscopic observations in the mid-infrared indicated the
presence of a \refmark{large} ($\sim$ 5.6 au) dusty viscous disk surrounding the
sgB[e] star in which the compact object is possibly enshrouded, thereby
giving the high $\mathrm{N_H}$ \citep{chat2012}. 

Indications of an orbital periodicity in the \textit{Swift} BAT lightcurve
were reported in 2009 by \citet{jain2009}. The authors found a periodicity
of $\sim$ 80 days and a single large peak in the orbital intensity profile
along with smaller secondary peaks. However, no periodicities or pulsations at smaller
time-scales have been reported for this source, despite the fact that
multiple searches have been made
using data from different observatories \citep{walt2003,barr2009}.

We investigated persistence of the orbital intensity modulation and
the source's flaring nature with additional data from 
\textit{Swift} BAT and \textit{INTEGRAL} ISGRI. We also examined archival 
data from the \textit{XMM-Newton}, \textit{Swift}, \textit{NuSTAR} and 
\textit{ASCA} observatories to search for pulsations and to check for
orbit phase \refmark{dependent} variations in the spectrum. We present
the results of this investigation in the following sections. In \S
\ref{sec:obs} we summarize the observations used and the steps
we followed to reduce the data from various observatories. We also list the
main results that we obtained in this section. In \S \ref{sec:dis} 
we discuss the relevance of these results and try to piece together an
overall picture of the binary system.

\section{Observations and analysis}\label{sec:obs}
\begin{figure}
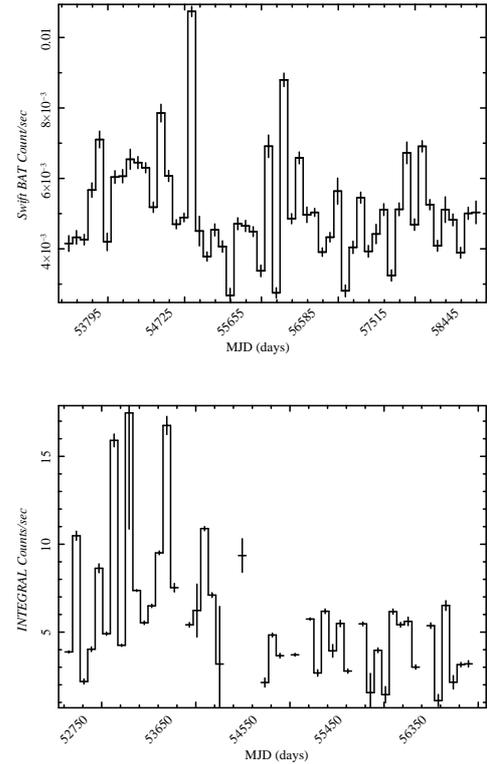

  \centering
  \includegraphics[width=0.4\textwidth]{batlicu.ps} \\
  \includegraphics[width=0.4\textwidth]{intlicu.ps}
  \caption{Lightcurves from \textit{Swift} BAT (top panel) and
  \textit{INTEGRAL} ISGRI (bottom panel) binned at \refmark{$\sim 7\times 10^6$} s.
  \textit{INTEGRAL} has lot of gaps in the data as compared to
  \textit{Swift} BAT. }
  \label{fig:licu}
\end{figure}

In-order to examine the intensity
variations in this source, we divided our analysis into two parts. The first part relates to the long term monitoring observations
and the second part relates to individual pointed mode observations. 
The long-term data was obtained from two sources. These were the
one day averaged lightcurve of \textit{Swift} BAT 
\footnote{\label{ft:bat}\url{https://swift.gsfc.nasa.gov/results/transients/}}
and the pre-reduced one hour bin-size lightcurves from \textit{INTEGRAL} ISGRI
\footnote{\url{http://www.isdc.unige.ch/integral/heavens}}. These two
lightcurves are shown in Figure \ref{fig:licu}\refmark{, with a bin-size equal
to the orbital period (discussed later). } 

The pointed observations were obtained from the HEASARC data
archive\footnote{\url{https://heasarc.gsfc.nasa.gov/docs/archive.html}}. Table
\ref{tab:obs} lists these observations. 
As the archival pointed observations were not targeted at
studying variations with orbital phase, we find that certain orbital phases
are not covered. As we note later, the absence of pointed observations
in some orbital phases limits the conclusions we can draw from these
observations.

\begin{table*}
  \centering
  \caption{Archival pointed mode observations used for studying spectral variations with
  orbital phase.}
  \label{tab:obs}
  \begin{threeparttable}
  \begin{tabular}{cccccp{4cm}}
	\hline
	 \textbf{Date (UTC)} & \textbf{Obs-id} &
	 \textbf{Observatory} & \textbf{Orb. phase}\tnote{*} & \textbf{Exposure
	 (ks)}
	 & \textbf{Remarks} \\
	\hline
	 1994-09-03 &  92000180   & \textit{ASCA} & 0.57 & 10.78 &\\
	 2003-02-10 &  0154750401  & \textit{XMM-Newton} & 0.05 & 26.98 &\\
	 2004-02-18 &  0201000201  & \textit{XMM-Newton} & 0.70 & 21.92 &\\
	 2004-03-20 &  0201000301  & \textit{XMM-Newton} & 0.10 & 25.52 &\\
     2004-08-20 &  0201000401  & \textit{XMM-Newton} & 0.00 & 21.91 &\\
	 2006-08-14 &  401094010  & \textit{Suzaku} & 0.66 & 97.25 &\\
	 2007-05-04 &  00035053003 & \textit{Swift} & 0.32 & 6.9 &Longest duration
	 \textit{Swift} observation with 57 photons in XRT \\
     2014-08-22 &  0742270201  & \textit{XMM-Newton} &  0.63 & 94.0&\\
     2014-08-22 &  30001006002  & \textit{NuSTAR} & 0.63 & 56.57 &Taken
	simultaneously with \textit{XMM} obs 0742270201 \\ 
	\hline
  \end{tabular}
  \begin{tablenotes}
    \item[*] Reference epoch is MJD 53477.577 and
	  period is 6920742.0 s. Phase is of mid-observation. See text for details.
  \end{tablenotes}
\end{threeparttable}
\end{table*}

The pointed mode observations were reduced using the standard observatory
pipelines made available by each of the X-ray observatories. Data from the
EPIC-pn instrument onboard \textit{XMM-Newton} was used
for extracting spectral information. This was done using the XMM software
suite \texttt{SAS v15.0.0}, with standard reduction techniques as specified
in the XMM-SAS user guide
\footnote{\burl{http://xmm-tools.cosmos.esa.int/external/xmm\_user\_support/documentation/sas\_usg/USG/SASUSG.html}}.

\begin{figure*}
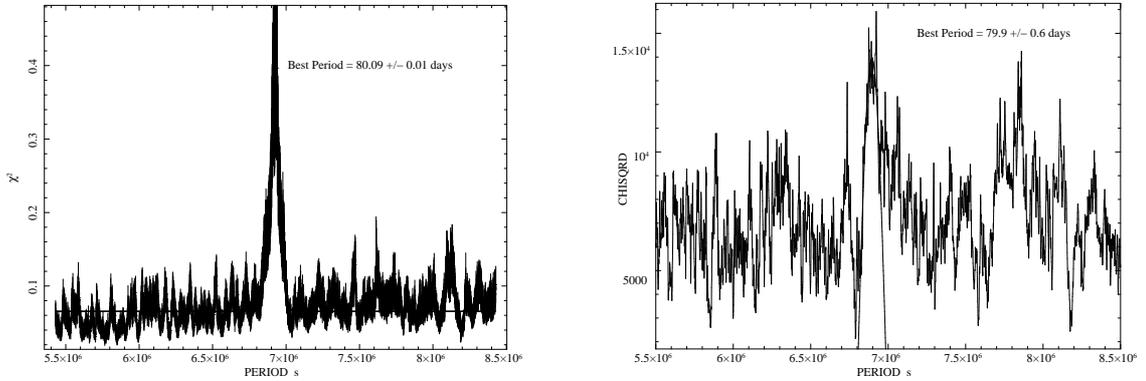

  \centering
  \includegraphics[width=0.45\textwidth]{batefs.ps}
  \includegraphics[width=0.45\textwidth]{intefs.ps}
  \caption{Results from the period search using \textit{Swift} BAT lightcurve
  (left panel) and \textit{INTEGRAL} lightcurve (right panel). Figure
plotted without error bars for clarity. The \refmark{most likely value of the
period is found by fitting a Gaussian curve to the peak in chi-square
distribution obtained from the period search.}}
  \label{fig:efs}
\end{figure*}

We used data from both the focal plane modules (FPMA and FPMB)  of the
\textit{NuSTAR} observatory. In order to reduce \textit{NuSTAR} data, the
software tools \texttt{nuproducts} and \texttt{nupipeline} were used from
the software suite \texttt{HEASoft v6.20}. The calibration files for
\textit{NuSTAR} were updated till Jan 2017. The data reduction was done as
per the instructions in the NuSTAR data analysis software user guide
\footnote{\url{http://heasarc.gsfc.nasa.gov/docs/nustar/analysis/nustar\_swguide.pdf}}.
We also made use of data from the XRT instrument onboard \textit{Swift}, XIS
and PIN instruments onboard \textit{Suzaku} and GIS instrument onboard
\textit{ASCA} \refmark{to obtain} the X-ray spectrum. The data
reduction for these instruments was also done using tools provided in the
\texttt{HEASoft v6.20} package with calibration files updated till Jan 2017. 
To fit the spectrum obtained from the reduced data, we used \texttt{XSPEC}, a
spectral fitting software package in \texttt{HEASoft}.

\subsection{Long-term Observations} 
We checked for signatures of the orbital modulation in both 
the \textit{INTEGRAL} ISGRI and \textit{Swift} BAT lightcurves. 
We used both the daily \refmark{averaged (DA) and the satellite orbit averaged
(SOA) lightcurves from BAT} for this analysis. However, the 
BAT imaging process can cause additional systematic errors to be present in
the data due to multiple possible reasons \citep[see][for
details]{krim2013}. Before carrying out a detailed analysis, we excluded data 
with error bars larger than $10^3$ times the minimum error value in the BAT
lightcurve.

\begin{figure*}
  \centering
  \includegraphics[width=0.69\textwidth]{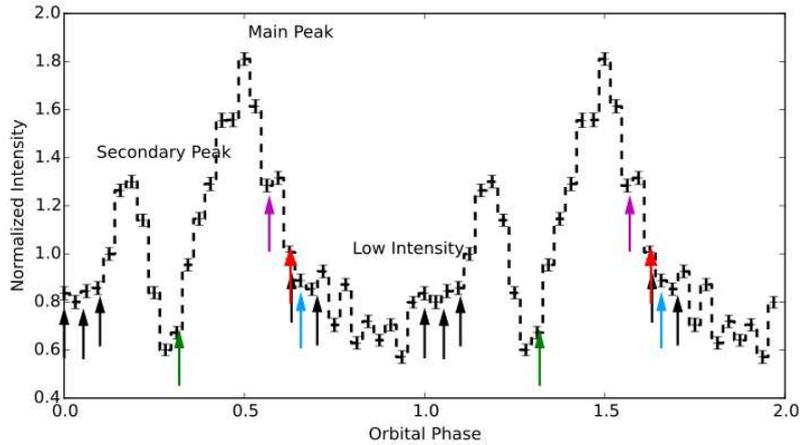}
  \caption{Orbital intensity profile as obtained from \refmark{\textit{Swift}
BAT lightcurve} showing orbital intensity modulations as a function of orbital
phase. The figure also shows the phase of each of the pointed mode
observations  with \textit{XMM} in black, \textit{Swift} in
green, \textit{NuSTAR} in red, \textit{ASCA} in magenta and \textit{Suzaku}
in blue arrows. See online for a color version of this figure.}
  \label{fig:fol}
\end{figure*}

 We searched for signatures of an
orbital period in the BAT DA lightcurves and in the ISGRI lightcurves.
This was done by using the tool
\texttt{efsearch} on the long-term lightcurves. \texttt{efsearch} searches
for periodicity by folding the lightcurve at different test periods and
comparing the folded profile against a constant non-varying profile
\citep{leah1983}. Given the previous detection of an orbital period at
$\sim$ 80 days, we searched for a periodicity in the period range from 60
days to 100 days. The results so obtained are shown in Figure \ref{fig:efs}.
As seen in the figure, there is a clear signature of periodicity
in both the lightcurves at $\sim$ 80 days. Since the
\textit{INTEGRAL} data has multiple long data gaps (100 - 500 d) between two
pointings of the galactic centre region, where this source resides, the
periodicity in ISGRI lightcurve is not as significant as in the BAT
lightcurve. The strong signature of periodicity in the BAT lightcurve
and an indication of a periodicity at the same period in
the \textit{INTEGRAL} lightcurve confirms \refmark{ the periodicity
reported} in \citet{jain2009} and marks it as an orbital
period. This period of $\sim$ 80 days is the highest reported orbital period in the small
sample of highly obscured sgHMXBs that are known currently. For all further
analysis, we used the
period \refmark{detected with \textit{Swift} BAT} at 80.09 $\pm$ 0.012 days.

The orbital period folded profile with folding epoch taken at MJD 53477.577 and
32 phase \refmark{bins per period} is shown in Figure \ref{fig:fol}. The epoch for folding is
taken such that the highest intensity appears at orbital phase 0.5. From the
figure it is clear that the intensity goes high in two phase
sections of the orbit, namely the phase between 0.10-0.25 and the phase
between 0.35-0.65. Now on, we shall refer to the brightest section as the Main Peak, the smaller hump
as the Secondary Peak and the rest as Low Intensity
phases, as is demarcated in Figure \ref{fig:fol}. This enables us to examine
the characteristics of each of these \refmark{phases} separately in order to
understand the reason behind such intensity variations. 

To further examine this dual peaked intensity variation, we took a closer
look at the long-term lightcurves and the X-ray spectra from each of the
pointed mode observations. The results of these are described in \S
\ref{ss:poOb}.

\subsubsection{Orbital timing analysis}
In order to see if the two peaks are persistently present in all orbits or if they are a
result of intermittent flares in some orbits \citep[as was reported in a
short duration of BAT lightcurves in ][]{jain2009}, we divided the long-term
\textit{Swift} BAT lightcurve into individual orbits (of 80.09 days). The BAT data
obtained for $\sim$ 12 years since 2005, gave us about 54 orbits for this
source. \citet{krim2010} reported two of the consecutively seen flares
from \textit{Swift} BAT to be separated by three times the reported orbital
period. We examined the data closely to see if such orbital phase dependence
of flares was true for all flares. 

Using the same criteria as in \citet{krim2010}, viz. flux greater than
130 mCrab ( 0.029 cts/s/cm$^2$) and significance greater than 10, we picked
all data points which can be classified to be flaring in the BAT SOA
data. In addition, we used the conditions \texttt{DATA\_FLAG ==  0}
and \texttt{DITHER\_FLAG == 0}\footnotemark[1]. 
Of the $\sim$ 50000 points in the SOA data, only 77 points were seen to
satisfy this criteria. If we relax the condition for dithering to be
present, we get 124 data points. Since this analysis was done on the SOA
data, where contributions to the systematic error due to non-dithering mode
observations are relatively less, we consider all 124 data points to be
valid flares.

\begin{figure}
  \centering
  \includegraphics[width = 0.47\textwidth]{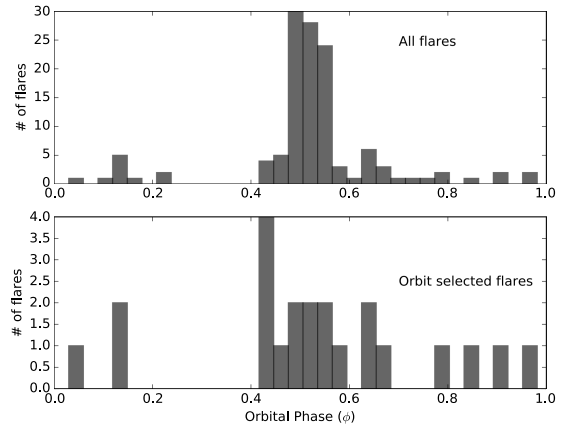} 
  \caption{Phase distribution of flares in the orbit. Top panel shows 
	orbital phase of all flares. Bottom panel shows orbit selected flares.
	See text for details.}
  \label{fig:outbat}
\end{figure}

\begin{figure*}
  \centering
  \includegraphics[width = 0.44\textwidth]{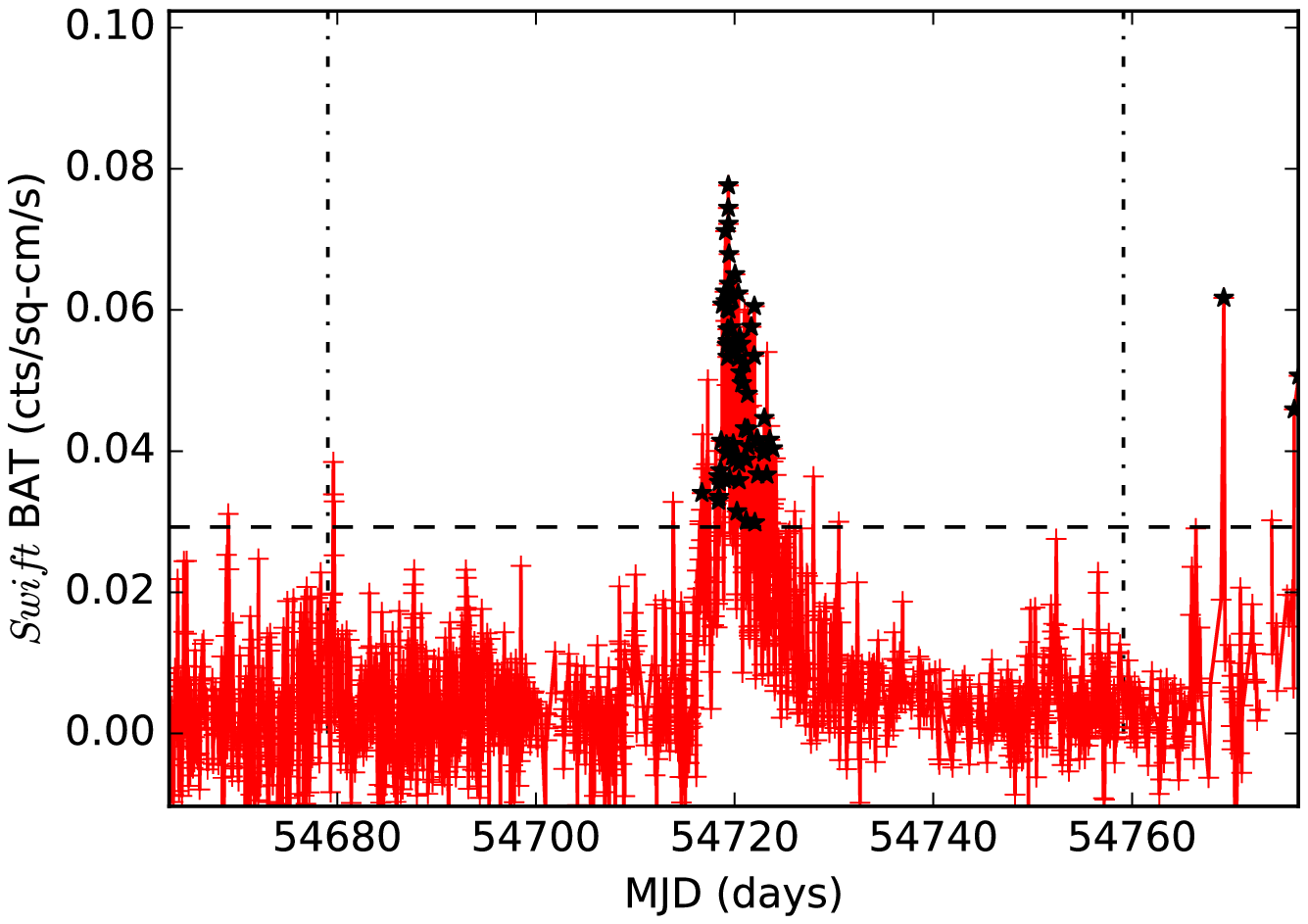}
  \includegraphics[width = 0.44\textwidth]{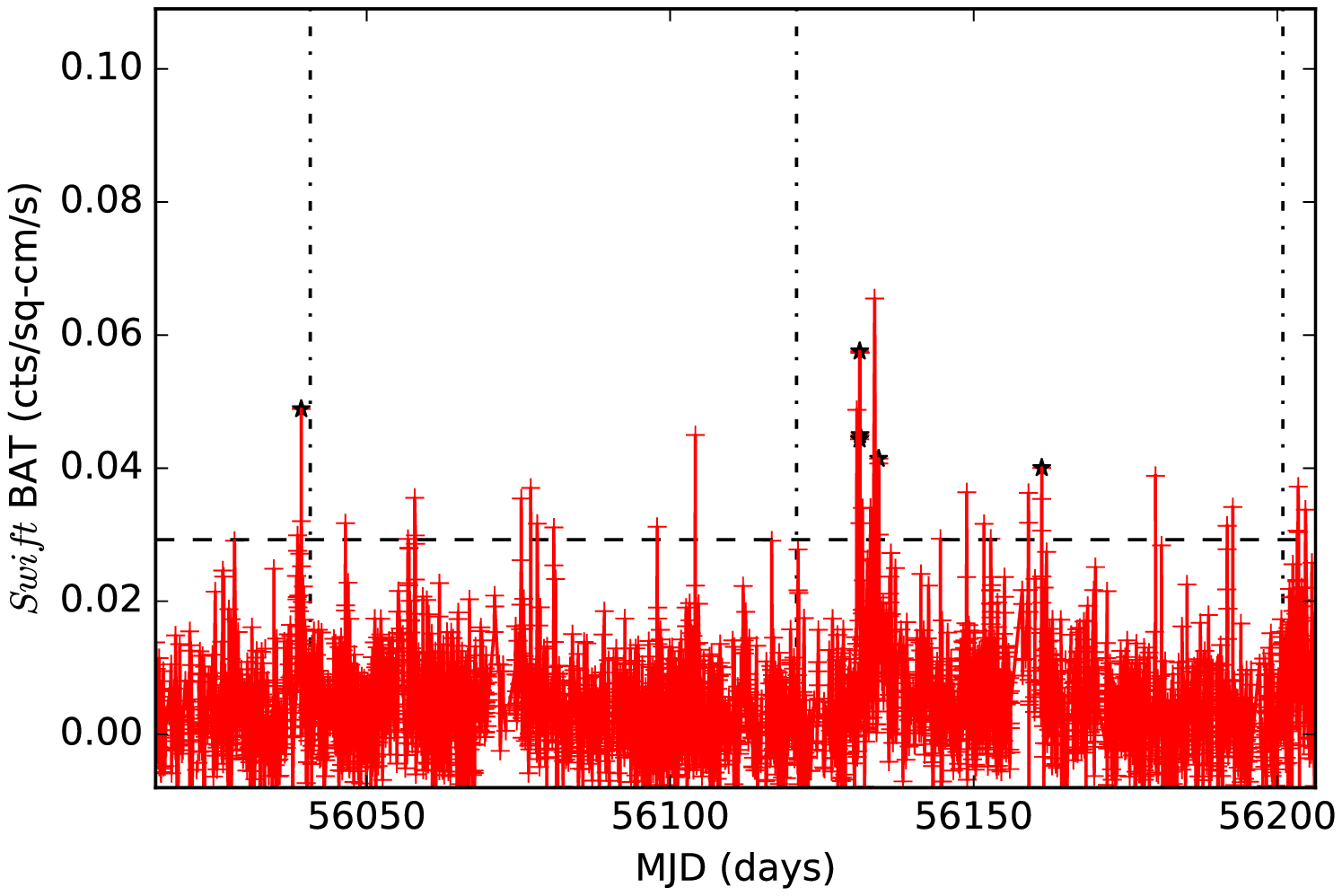}
  \caption{Figure shows the BAT SOA lightcurve for a \refmark{few orbits of data}
	with flares detected in \textit{Swift} BAT SOA. Left panel shows flares in a
	major outburst in 2008. Right panel shows some of the data points detected in
the Secondary Peak in 2012. In both figures, the
horizontal line corresponds to the 130 mCrab threshold and the vertical
lines correspond to the ephemeris at phase zero of each orbit. SOA data points
selected to be in a flaring state are marked in black. See online text for a color version of this
figure.}
  \label{fig:outbat2}
\end{figure*}

Using the same orbital period and epoch used to fold the BAT lightcurves, we tried to see
the orbital phase at which the flares occur. Figure \ref{fig:outbat} shows the phase 
distribution of these flares. As is immediately evident from this figure, 
most (99 of 124) of the flares occur in the phase corresponding to the Main Peak. 
We also \refmark{note that} quite a few (8) flares exist in the phase which 
corresponds to the Secondary Peak. It is to be noted that a majority (64) of the 
points in flares in the Main Peak belong to one outburst (consisting of
multiple consecutive flaring points ) in September 2008, as seen in 
left panel of Figure \ref{fig:outbat2}. We checked to see if all other flares
happen in the Main Peak by looking at orbit selected flares.

This was done by taking only one representative point for every orbit by
recording the phase of the flare with the highest flux in that orbit.
Making a histogram of this (bottom panel of Figure \ref{fig:outbat}) still
shows that a majority (13 of 22) of the orbits have flares in
Main peak. A further 3 points lie just after the Main
peak (0.65-0.8) and 2 points lie in the Secondary Peak. This
accounts for 18 of 22 orbits showing flares, 
thereby indicating a strong orbital phase preference for these flares.
However, it is interesting to note that a few orbits (4 of 22) show
flares which are not in either of the two Peaks.

\begin{figure*}
  \centering
  \includegraphics[width=0.45\textwidth]{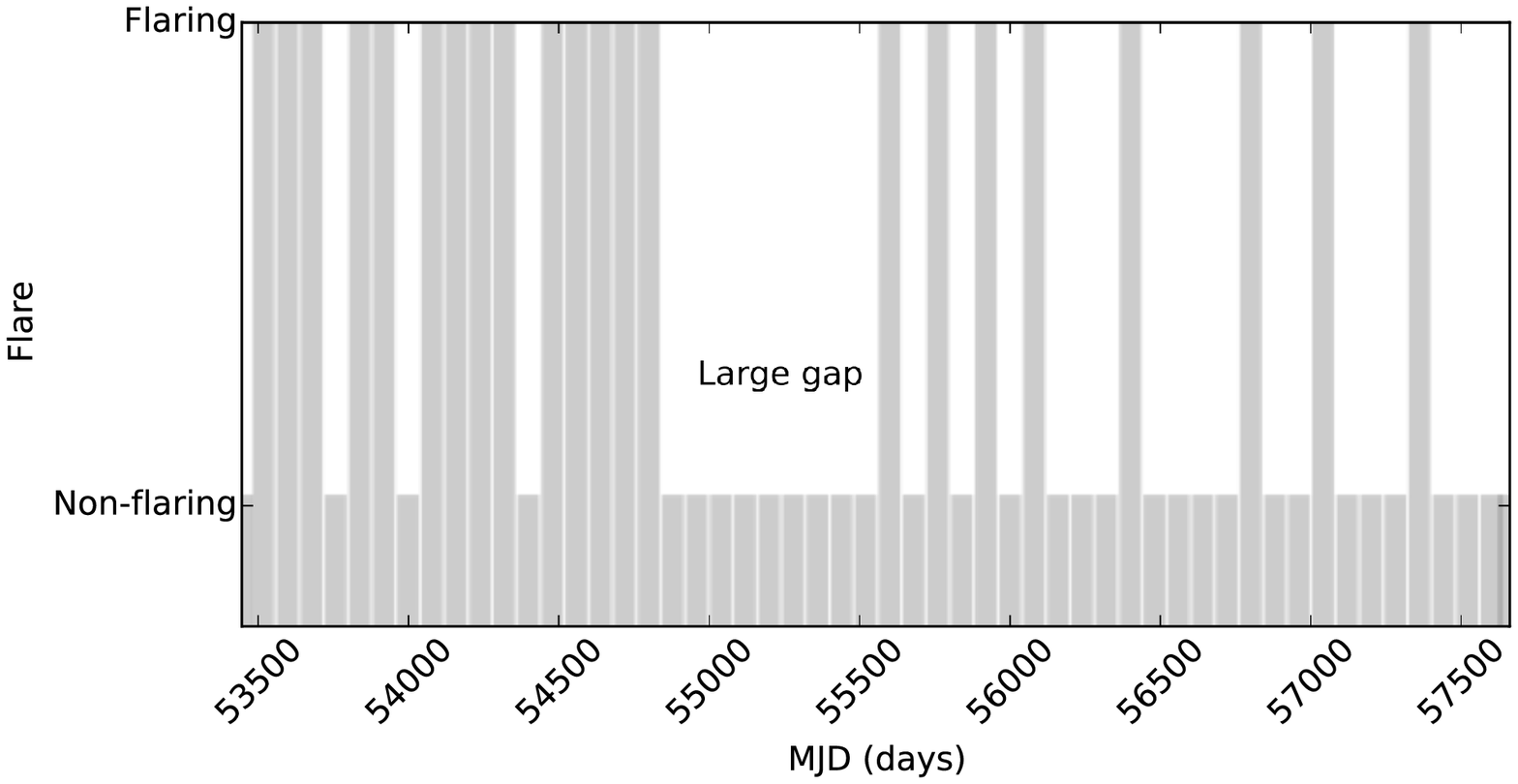}
  \includegraphics[width=0.45\textwidth]{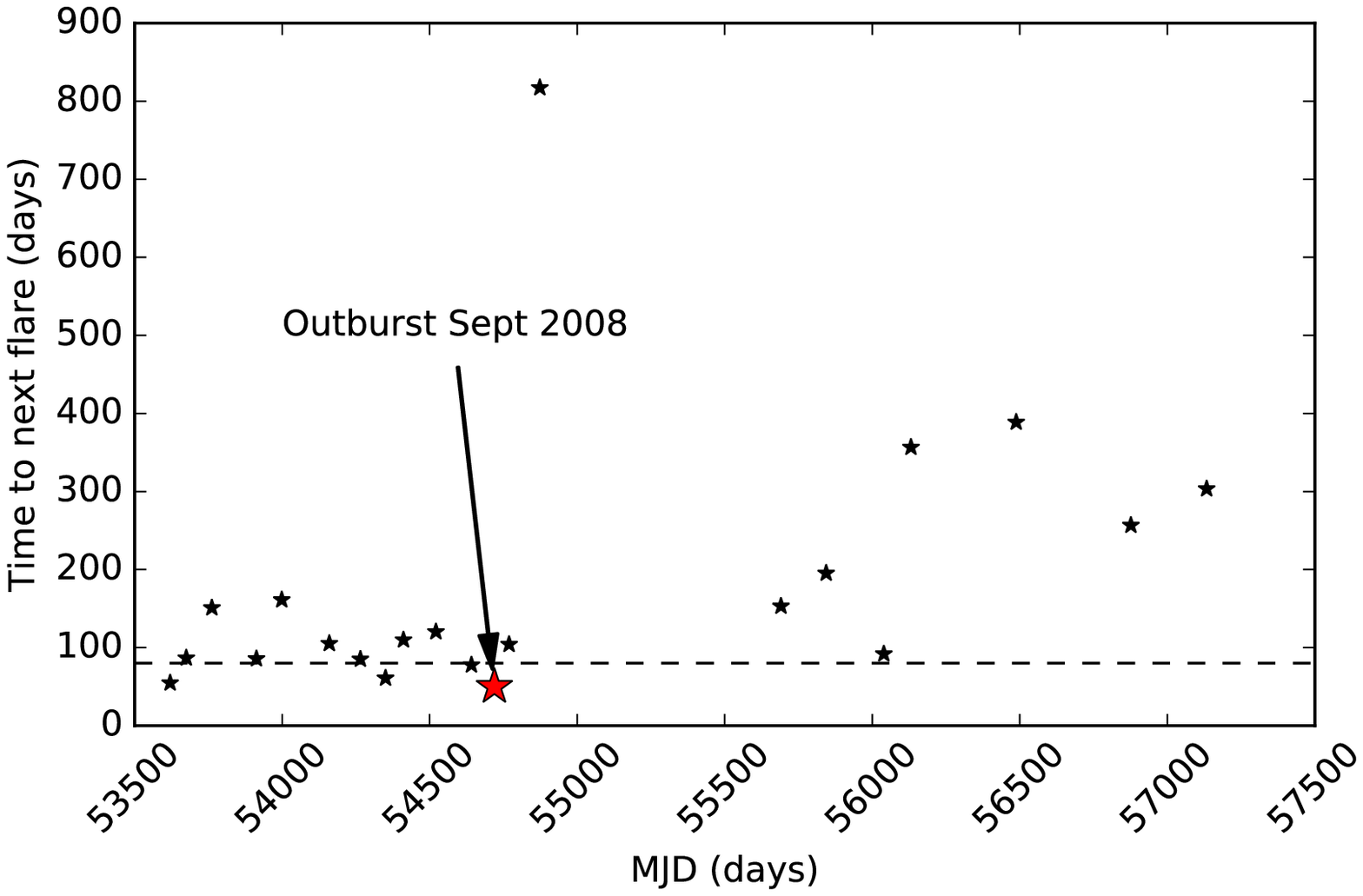}
  \caption{Left panel shows orbits in which flares were detected. Almost
  every orbit had a detected flare before a large gap without flares
  was seen starting Feb 2009 (orbit number 18).
Right panel shows time to next flare. As seen after MJD 54890 (Feb 2009)
the time to \refmark{next} flare shows a significant increase as compared to pre 2009
flares. }
  \label{fig:burs}
\end{figure*}

We tried to see if there is a recurrence rate to the occurrence of flares
by noting the orbits in which the flares occurred. \refmark{ However, as} plotted in 
left panel of Figure \ref{fig:burs} there seems to be no periodic pattern to the 
occurrence of these orbits. The only noticeable change seems to be the fact 
that the source was in a frequent flaring mode before Feb 2009. 
This is also seen from the right panel of Figure \ref{fig:burs} where the time to next 
flare is plotted. From these plots, we note that there is a $\sim$ 10 orbit gap after the
detected flare in Feb 2009. Post this large gap, the time to next flare 
shows a sporadic distribution, with values much larger than those before
Feb 2009. We note that in Sep 2008, two orbits before the large gap with no
flares, a major outburst is seen in the \textit{Swift} BAT data (also see left
panel of Figure \ref{fig:outbat2}). This behaviour seems similar to disk-fed systems where the time to 
\refmark{next outburst} is directly proportional to the luminosity of the outburst, suggesting
some kind of disruption and reformation of the accreting material in the disk
\citep[see the case of dwarf novae as explained by the Disk instability model in][and
references therein]{osa1996}.

\begin{figure*}
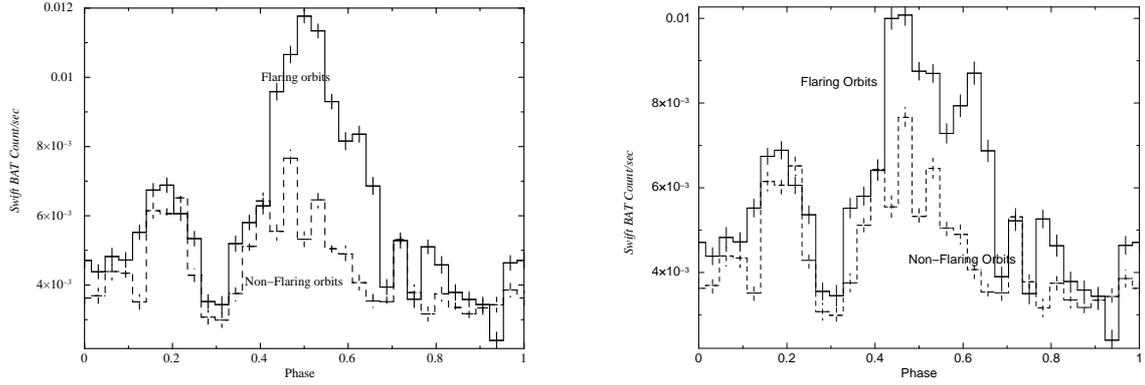

  \centering
  \includegraphics[width=0.45\textwidth]{bnborb.ps}
  \includegraphics[width=0.45\textwidth]{bnborb2.ps}
  \caption{\refmark{Figure compares the folded profiles of orbits with flaring
  observations to orbits without any flares. The right panel constructs the
folded profiles by ignoring data points from the large outburst of Sep. 2008}}
  \label{fig:proff}
\end{figure*}

\refmark{\subsubsection{Orbit folded profiles}
The orbital phase preference for occurrence of flares is seen in the previous section.
However, the source lightcurves also seem to show the orbital modulation
in the absence of flares. This is seen in Figure \ref{fig:proff}.
For making this figure, the 22 orbits which had at-least one flaring observation were
folded together and compared with the folded profile of the remaining 32 orbits
which had no flaring observations. The figure indicates that non-flaring orbits also have enhanced
emission at the phases corresponding to the Main and Secondary peaks, with the
folded amplitudes in the two peaks being comparable. However, flaring orbits seem
to have a much higher amplitude in the Main peak. We note that the large outburst of Sep 2008 
biases these folded profiles. Therefore, we also compared the folded profiles 
by neglecting data from this outburst as seen in the right panel of Figure
\ref{fig:proff}. The Main peak continues to have a higher amplitude in the
flaring orbits as compared to non-flaring orbits. This indicates that the
preference for flares to occur in the phases corresponding to the Main peak
increases its amplitude as seen in the folded profile of the entire lightcurve, 
while the Secondary peak corresponds to a phase of enhanced emission in almost
all orbits.}

\begin{figure*}
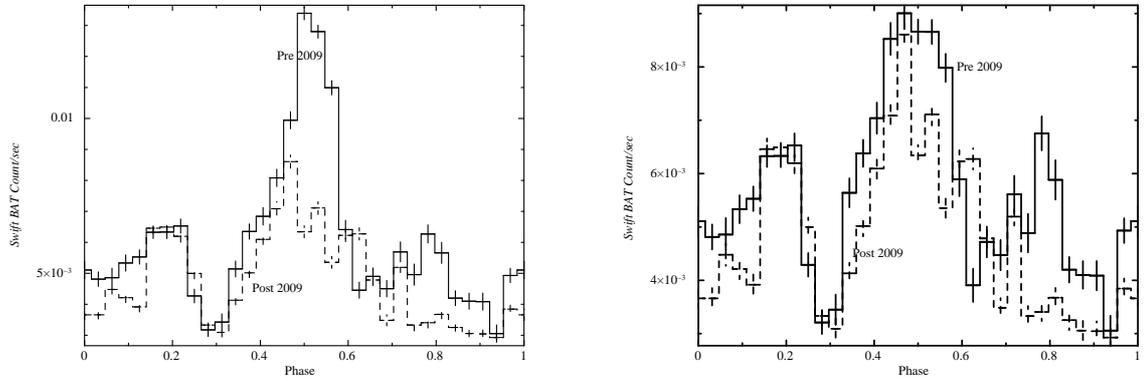

  \centering
  \includegraphics[width=0.45\textwidth]{prepos.ps}
  \includegraphics[width=0.45\textwidth]{prepos2.ps}
  \caption{\refmark{Figure compares the folded profile before the large gap
	starting Feb 2009 to the profile after the large gap. The right panel
	constructs these profiles by ignoring data points from the large outburst of
  Sep. 2008}}
  \label{fig:proff2}
\end{figure*}

\refmark{We also tried to see if the folded profile changed before and after the large
gap starting Feb 2009. The lightcurves corresponding to these time periods were
folded separately, with the resultant profiles plotted in Figure
\ref{fig:proff2}. We folded these profiles with and
without the outburst of Sep. 2008, similar to Figure \ref{fig:proff}. We find
that the folded profiles do not change much before and after the
large gap if the Sep. 2008 outburst is not considered.}

\subsection{Pointed Observations}\label{ss:poOb}
The pointed mode observations \refmark{of IGR J16318-4848 too} show intensity variations over the course
of a single observation \citep{matt2003}. We tried to
check if these variations too have any signatures of periodicity. We also
examined how these variations differ across different X-ray energy
bands. Finally, we checked to see how these short-term variations link with
the orbital variations in the source. The following paragraphs detail the
results of this investigation.

\subsubsection{Search for pulsations}\label{ss:puls}
Most HMXB systems have a neutron star as the compact object
\citep[see][for an exception]{cas:2014}. \citet{walt2003} reported absence
of pulsations in \refmark{IGR J16318-4848 using} \textit{XMM} observations, with a marginal signature
of a QPO at 0.15 Hz. \citet{barr2009} noted that even the harder X-rays as
seen by \textit{Suzaku} gave no pulsation signatures in \refmark{ this source in
the range from} 1s to 10 ks. 

\begin{figure*}
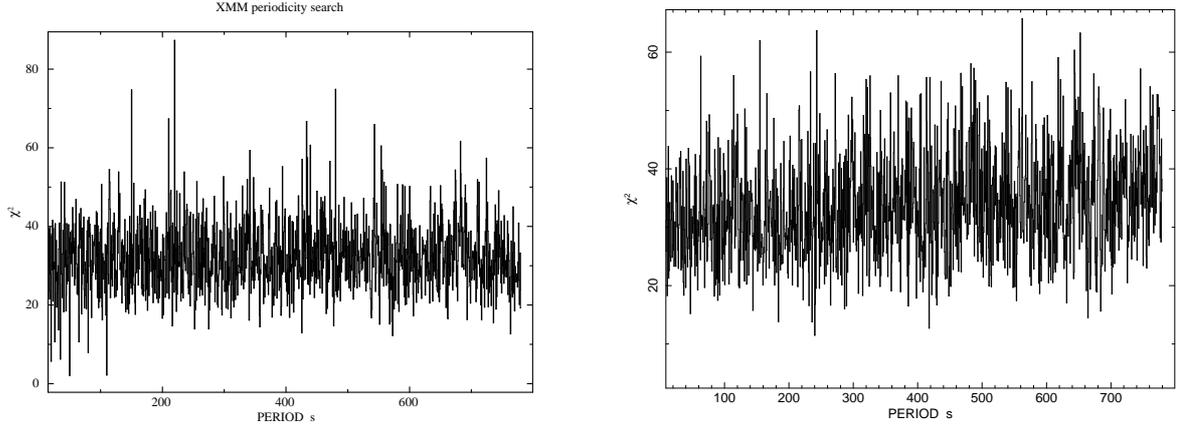

  \centering
  \includegraphics[width=0.45\textwidth]{xmmefs.ps}
  \includegraphics[width=0.47\textwidth]{nuefs.ps}
  \caption{Results of pulsation search in \textit{XMM} (left panel) and
  \textit{NuSTAR} (right panel) data from 0.5s to 1ks. The non-detection
  of pulsations enabled us to put an upper limit \refmark{ of $\sim$ 1\% for
	\textit{XMM} and $\sim$ 4\% for \textit{NuSTAR} on the pulse fraction of 
  any un-detected pulsation.}}
  \label{fig:puls}
\end{figure*}

\refmark{We searched for pulsations in} the \textit{NuSTAR} lightcurve
and the simultaneously taken \textit{XMM} lightcurve for pulsation signatures.
This \textit{XMM} data was taken
in the small window mode, enabling a time resolution of 5.7 ms. The
lightcurves for both \textit{XMM} and \textit{NuSTAR} were barycentered
before the search was made. 
Figure \ref{fig:puls} shows the results of \refmark{the period search}. As seen in figure, no
pulsations are detected either in the soft ( $<$ 10 keV) or hard ( upto 70
keV) bands, with an upper limit on pulse fraction of $\sim$ 1\% in
\textit{XMM} and $\sim$ 4\% in \textit{NuSTAR}. The QPO feature at $\sim$ 0.15 Hz was 
also not seen in the power density spectrum from these lightcurves. 
We do note that \texttt{efsearch} gives
high $\chi^2$ values for periods greater than 2000 s in both the \textit{XMM}
and \textit{NuSTAR} data. We believe that this could be due to \refmark{an inherent
slow kilosecond scale} variability in the source as pointed out in the next section \S \ref{ss:varspec}.

\subsubsection{Intensity and spectral variations at smaller timescales -
The \textit{XMM} -\textit{NuSTAR} simultaneous observation}\label{ss:varspec}
The source \refmark{exhibits} a significant variation
of intensity over the course of a single pointed observation. The
source intensity often changes by an order of magnitude over a few
kiloseconds \citep[see the first \textit{XMM} observations reported in][]{matt2003}. 
This variation is reflected in both the power density spectrum (with
high power in the sub mHz region) and \texttt{efsearch} results.

\begin{figure*}
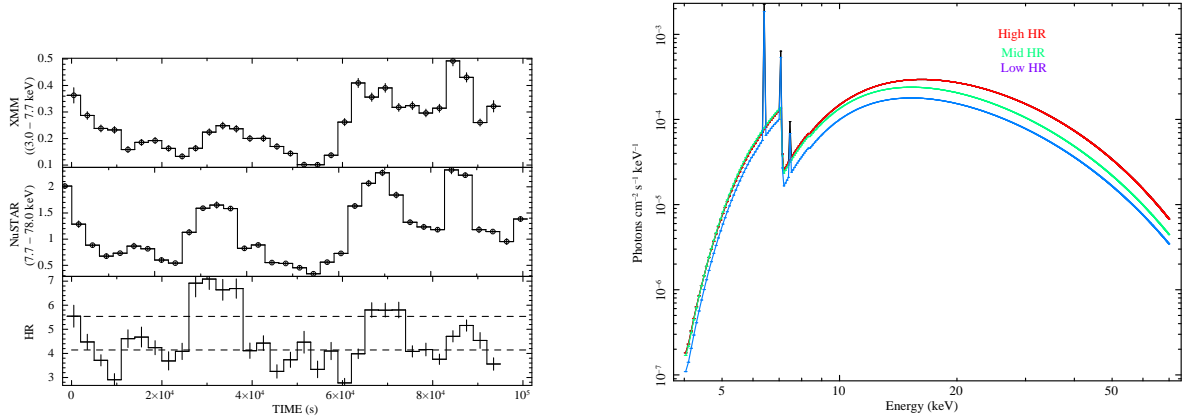

  \centering
  \includegraphics[width=0.45\textwidth]{xmmnu.ps}
  \includegraphics[width=0.47\textwidth]{spechr.ps}
  \caption{\refmark{Hardness ratio (HR)} and spectral variations over a pointed observation. Left panel
	shows \refmark{HR variations and the \textit{XMM} and \textit{NuSTAR} lightcurves in
  the observation period}.
	Horizontal lines show levels used to create the HR resolved
spectra. Right panel shows the best fit models indicating the spectral changes as a 
	function of change in HR value for three HR levels (red for high,
	\refmark{green} for mid
  and blue for low).See online text for a color version of this
figure.}
  \label{fig:hrvar}
\end{figure*}

We tried to see if these intensity variations over the course of a pointed observation
lead to hardness ratio and spectral changes. 
The left panel of Figure \ref{fig:hrvar} plots the lightcurves in the soft
\textit{XMM} (3.0 - 7.7 keV) and hard \textit{NuSTAR} (7.7 - 78 keV) bands.
The bands were chosen such that the line energies and the most of the continuum
fall in separate bands. The variations
in both these bands seem well correlated as indeed was observed by
\citet{matt2003} and \citet{barr2009}. \citet{matt2003} also noted little to
no change in 
the spectral parameters (other than overall normalization) between the observations at
different luminosity levels. The hardness ratio (HR) plot of \refmark{the
\textit{Suzaku} instruments \citep[Figure 5 of ][]{barr2009} shows
changes in HR values over the \textit{Suzaku} observation. However, it is
difficult to comment on these changes and their relation to change in X-ray
flux. The HR plot} obtained by the more sensitive observation using simultaneous
stares by \textit{XMM} and \textit{NuSTAR} shows a clear increase in
HR values (by a factor of 2) with increase in luminosity. This indicates a change
in spectral parameters causing increase in source hardness with increase in
source luminosity. We examined the spectra of  \textit{XMM / NuSTAR} 
observations as a function of the HR values to \refmark{investigate} this.

\begin{table*}
  \centering
  \caption{Absorption, line and continuum parameters as a function of
	hardness ratio changes in the \textit{XMM/NuSTAR} observation}
  \begin{tabular}{c|cccccc}
	\hline
	\textbf{Obs} & \textbf{Nh $(\times 10^{22})$} & 
	\textbf{Fe-K$\alpha$ EW (keV)} & \textbf{$\Gamma$} & \textbf{Ecut} &
	\textbf{Efol} & \textbf{$\chi^2$/dof} \\
	\hline
	HR high & 116.68$^{-1.61}_{+1.67}$ & 1.86$^{-0.12}_{+0.12}$ &
	-0.42$^{-0.09}_{+0.10}$ & 7.1$_{-0.19}^{+0.66}$ & 11.3$_{-0.13}^{+0.12}$ & 421.54/387\\
	HR mid & 123.95$^{-1.55}_{+1.58}$ & 1.55$^{-0.09}_{+0.09}$ &
	0.07$^{-0.10}_{+0.10}$ & 7.0$_{-0.29}^{+0.18}$ & 12.6$_{-0.23}^{+0.11}$ & 460.45/401 \\
	HR low & 127.46$^{-1.27}_{+1.26}$ & 1.93$^{-0.19}_{+0.19}$ &
	0.15$^{-0.10}_{+0.11}$ & 6.7$_{-0.19}^{+0.25}$ & 13.2$_{-0.19}^{+0.22}$ &
320.26 / 348 \\
\refmark{Avg Spec} & \refmark{124.10$^{-0.63}_{+0.59}$} & \refmark{1.80$^{-0.05}_{+0.05}$} &
\refmark{-0.04$^{-0.05}_{+0.05}$} & \refmark{6.9$^{-0.14}_{+0.13}$} &
\refmark{12.4$^{-0.09}_{+0.09}$} & \refmark{1009.1/885} \\
	\hline
  \end{tabular}
  \label{tab:hrvar}
\end{table*}

\begin{figure}
  \centering
  \includegraphics[width=0.47\textwidth]{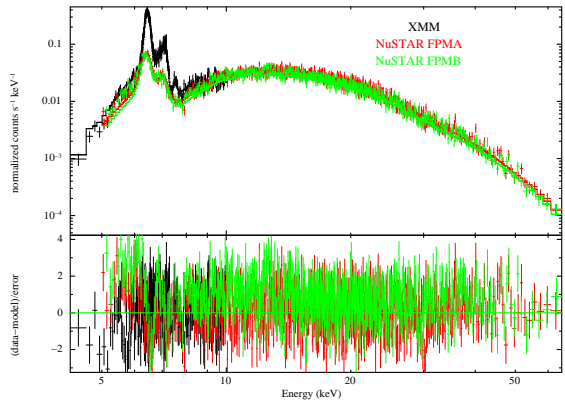}
  \caption{Spectrum averaged over entire \textit{XMM-NuSTAR} simultaneous
observation. See text for details. See online text for a color version of this
figure.}
  \label{fig:spec}
\end{figure}

The source spectra was obtained in three distinct HR levels denoted as high
(HR $>$ 5.5), mid (5.5 $>$ HR $>$ 4.0) and low (HR $<$ 4.0).  
The spectra were extracted as per the steps mentioned in \S \ref{sec:obs}. 
To fit the spectra, we first tried a \texttt{powerlaw} model with
absorption (\texttt{phabs}) and emission lines (\texttt{gaussian}). This
fit indicated the presence of a cut-off at higher energies. To model the
cut-off, we used the \texttt{highecut} model.We do note that we could have used 
\texttt{cutoffpl} instead. However, we found a large difference between the
photon index for data taken below 10 keV (using only \textit{XMM}) and for data 
taken till 70 keV (using \textit{XMM} and \textit{NuSTAR}) \citep[also
see][]{barr2009}. This difference is much lesser when the \texttt{highecut}
model is used, thereby enabling comparison of observations with broadband
data and observations with data up to 10 keV. Thus, the spectral model we use
consists of an absorption component (\texttt{phabs}), a continuum component
(\texttt{powerlaw} with \texttt{highecut}) and three line components (for
Fe-K$\alpha$, Fe-K$\beta$ and Ni-K$\alpha$). The average spectrum for the
entire observation fit by this model is shown in Figure \ref{fig:spec}.

The \refmark{best fit model incident spectrum} for the segments with different HR values are shown in right
panel of Figure \ref{fig:hrvar}. Table \ref{tab:hrvar} lists the parameters obtained from the fitting routine.
We note that the continuum parameters harden with increase in
source intensity as seen by \refmark{decrease in $\Gamma$ values.}

Thus, the variation of the broadband spectrum over a single observation of
$\sim$ 60 kilosecond duration is mainly related to change in
the photon index (and normalization) of the power-law component, which
causes related changes in the HR values and line equivalent widths. However,
with the other parameters being near constant over the course of an
observation, we use the average spectrum of the entire
observation in order to see if there are any systematic changes of the
spectral parameters with orbital phase.

\subsubsection{Spectral changes over the orbital duration}
As reported in Table \ref{tab:obs}, there are nine archival observations
distributed over various phases of the orbit of this source. Although more \textit{Swift}
observations were available, most of them contained too few integrated
counts (< 60) to be useful for spectral analysis, owing to the highly absorbed nature of
this source. Figure \ref{fig:fol} shows that none of the
observations cover either of the two peaks. All the pointed observations
lie in the low intensity phase of the orbital profile.

\begin{figure}
  \centering
  \includegraphics[width=0.48\textwidth]{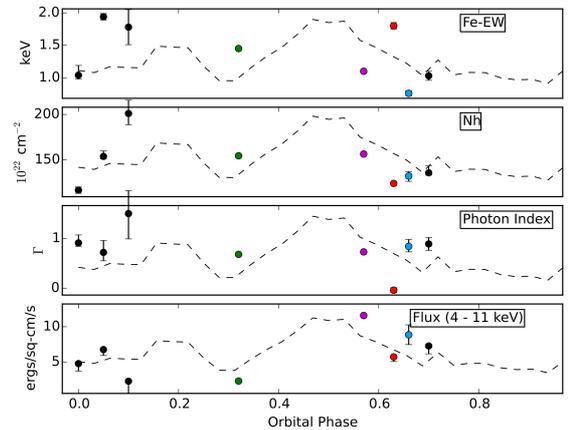}
  \caption{Variation of spectral parameters as a function of the orbital
  phase. The orbital profile is plotted alongside each figure for
  reference.\textit{Swift} and \textit{ASCA} data \refmark{do not have error bars plotted
  (See Table \ref{tab:spec}). The color scheme of Fig. \ref{fig:fol} is followed here to
distinguish data from different observatories. See online for a color version of
this figure.}}
  \label{fig:spectr}
\end{figure}

As seen from the bottom panel of Figure \ref{fig:spectr}, these 
pointed mode observations do seem to be following the luminosity trends 
of the BAT folded profile. Thus, we try and look for correlations between 
spectral changes in the pointed mode observations and the BAT folded profile flux. 
In order to look for spectral changes, we extracted the X-ray spectrum for each of the
pointed mode observations and modelled the spectrum as mentioned in the
\S \ref{ss:varspec}. For observations with data upto 10 keV (\textit{XMM},
\textit{Swift} and \textit{ASCA}), the high energy cut-off was not used while
modelling the data.

\refmark{ We do note that when we used a simple Gaussian model for the three
  lines, we ignore the effects of a Compton shoulder \citep[as used
  in][]{matt2003,ibar2007} or the presence of multiple lines of different
  ionization states \citep[as used in][]{barr2009}. As a result of this, although we cannot compare our
  results for line equivalent widths with these results, we get a simple way to
  compare all the archival observations in a consistent manner. Secondly, we
  also note that strong correlations exist between the power law index ($\Gamma$) and
  the column density ($\mathrm{N_H}$) \citep[see also Figure 1 of][]{matt2003} . 
  This can cause the multiple degenerate fits resulting into different local minima. 
  This degeneracy is broken with the availability of broadband data \citep[as seen in
  ][]{ibar2007,barr2009}. Excepting a couple of observations, all our
  datasets have spectrum only upto 10 keV. For these low energy observations, we try to
  consistently get the best fit value by checking for other local minima using
  the \texttt{steppar} command. It is noted that \citet{ibar2007} do model some
  of the data-sets with additional \textit{INTEGRAL} observations, thereby
  breaking the degeneracy. We use only the \textit{XMM} observations ( < 10 keV) 
  for these data-sets in order to keep the analysis consistent for all \textit{XMM}
observations. }

\begin{figure}
  \centering
  \includegraphics[width=0.45\textwidth]{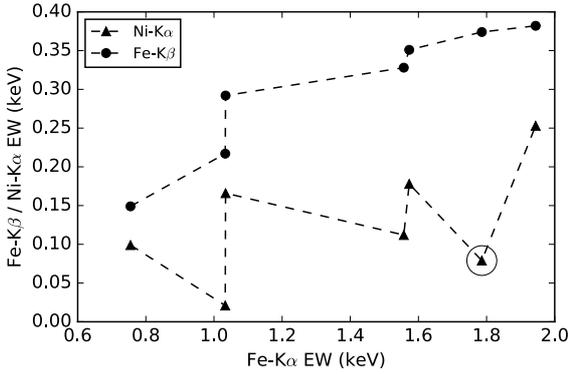}
  \caption{Correlated variations in the Fe-K$\alpha$ and Fe-K$\beta$ /
  Ni-K$\alpha$ observations. Circled point indicated observation with the
highest column density}
  \label{fig:linevar}
\end{figure}

To look for spectral changes we checked for change in one line parameter
(Fe-K$\alpha$ equivalent width), one absorption parameter (column density)
and one continuum parameter (photon index) as a function of
the orbital phase. These variations are plotted in Figure \ref{fig:spectr}
and tabulated in Table \ref{tab:spec}.
The fluxes of the two iron lines (Fe-K$\alpha$, Fe-K$\beta$) are
correlated with each other as expected. This is shown in the Fig.
\ref{fig:linevar}. However, the Nickel line variations (triangular points in
Figure \ref{fig:linevar}) do not seem to follow the Iron line variations. This,
though, could be a result of the very strong absorption present after the
Fe-K edge ( at $\sim$ 7.11 keV), which can cause the Nickel line to be below
detection limits for observations with a large column density. As the lines
are largely correlated, we just note the variation in Fe-K$\alpha$ with
change in orbital phase.

\begin{table*}
  \centering
  \caption{Spectral parameters obtained from fitting each of the pointed
  mode observations. Errors are 1 $\sigma$ values.}
  \begin{threeparttable}[b]
  \begin{tabular}{ccccccc}
	\hline
	\textbf{Observatory} & \textbf{Orb. phase} & \textbf{Nh $(\times
	  10^{22})$} & \textbf{Fe-K$\alpha$ EW (keV)} & \textbf{$\Gamma$} &
	  \textbf{Flux (4-11 keV) $(\times 10^{-12})$ ergs/sq-cm/s} \\
	  \hline
	  \textit{XMM} & 0.00 &  116.73$^{-3.50}_{+3.48}$ &
	  1.04$^{-0.06}_{+0.15}$ & 0.91$^{-0.07}_{+0.16}$ &
	  4.78$^{-1.06}_{+0.04}$ \\
	  \textit{XMM} & 0.05 &  153.83$^{-2.83}_{+5.99}$ &
	  1.94$^{-0.05}_{+0.05}$ & 0.72$^{-0.17}_{+.24}$ &
	  6.77$^{-0.85}_{+0.02}$ \\
	  \textit{XMM} & 0.10 &  201.05$^{-12.33}_{+14.69}$ &
	  1.78$^{-0.27}_{+0.27}$ & 1.50$^{-0.51}_{+0.46}$ &
	  2.29$^{-1.73}_{+0.02}$ \\
	  \textit{Swift}\tnote{*} & 0.32 & 154.50 & 1.45 & 0.68 & 2.31 \\
	  \textit{ASCA}\tnote{*} & 0.57 & 156.46 & 1.10 & 0.73 & 11.59 \\
	  \textit{XMM-NuSTAR} & 0.63 & 124.10$^{-0.63}_{+0.59}$ &
	  1.80$^{-0.05}_{+0.05}$ & -0.04$^{-0.05}_{+0.05}$ &
	  5.73$^{-0.66}_{+0.08}$ \\
	  \textit{Suzaku} & 0.66 & 132.31$^{-5.88}_{+4.75}$ &
	  0.76$^{-0.04}_{+0.04}$ & 0.84$^{-0.11}_{+0.14}$ &
	  8.87$^{-1.40}_{+1.40}$ \\
	  \textit{XMM} & 0.70 & 135.79$^{-2.45}_{+7.73}$ &
	  1.03$^{-0.07}_{+0.07}$ & 0.89$^{-0.13}_{+0.13}$ &
	  7.28$^{-1.12}_{+0.12}$ \\
	  \hline
  \end{tabular}
  \begin{tablenotes}
  \item[*] Data has very poor statistics, thereby giving very large errors in \textit{Swift} and
	\textit{ASCA} parameters. Only the central values are quoted here.
  \end{tablenotes}
  \end{threeparttable}
  \label{tab:spec}
\end{table*}

The absorption column density remains persistently high ($\mathrm{N_H} >
10^{24}$ cm$^{-2}$) through all phases of the orbit.
The iron line equivalent width too is seen to be fairly high ( $> 0.7$ keV) in
all phases of the orbit. The photon index $\Gamma$ is seen to vary across a single
observation, so drawing conclusions from its variation in the average spectrum
may not be useful. However, we do note that the obtained photon indices indicate a very
hard spectrum with values in the range (-0.1 to 1.5).

Given the limited coverage of different orbital phases by pointed mode
observations, it is difficult to draw conclusions from the trends in the
parameters. There seems to be an indication of a continuous rise 
in $\mathrm{N_H}$ values just before the Secondary Peak is seen. It would be 
interesting to have more observations to note the variations of these parameters 
in the Main and Secondary Peaks.

%The X-ray spectrum for \textit{XMM} observations were obtained directly from
%the Pipeline Processing System (PPS). For all \textit{XMM} observations other than
%obsid 0742270201, the files were processed in Dec 2012. For obsid
%0742270201, the files were processed in Sep 2014. We took the readily
%available EPIC PN spectrum, background and arf files made available in the
%PPS data. The response files were obtained from the canned response matrices
%made available online. For \textit{Swift} and \textit{NuSTAR} observations, 

\section{Discussion}\label{sec:dis}
From the timing analysis of this source, we see that there are two peaks in
the orbital folded profile. Based on the spectral and timing observations,
we try to speculate upon the reasons behind the formation of two such peaks in the
orbital folded profile. 

By selecting data points which are flaring in the BAT SOA light-curve, we see that
the Main Peak corresponds to a phase during 
which a large number of flares (factor of 10 or greater) are seen as was detected in
Sep 2008. The non-flaring data also has a signature of enhanced emission in
the Main and Secondary Peaks. However, the flaring data points have a
preference for orbital phase during which they occur. A possible picture of the accretion
process in this binary which could explain all of these is considered in the
following sub-section.

\subsection{Orbital picture}
\begin{figure}
  \centering
  \includegraphics[width=0.5\textwidth]{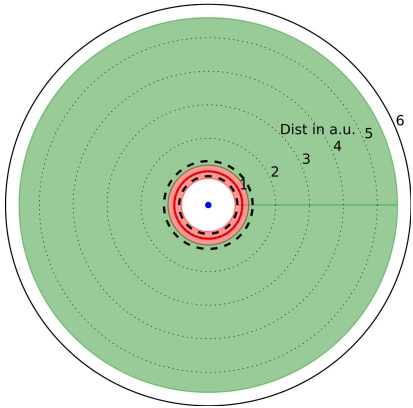}
  \caption{Illustration of the orbit placed in the companion's dust rim as
	seen from the orbit axis. Blue dot is the companion star, red annulus represents the dust rim and
  green annulus represents the warm dust disk \citep{chat2012}. Black dotted lines represent the minimum and maximum possible values of the
orbit semi-major axis. \refmark{The minimum semi-major axis for a neutron star
compact object lies at the dust rim and is not indicated separately here.}
Distances used to plot the diagram are given in text.
Figure assumes a circular orbit. See online text for a color version of this
figure.}
  \label{fig:orbit}
\end{figure}

Using mid-infrared spectroscopic observations, \citet{chat2012} inferred
\refmark{IGR J16318-4848 to be a} near edge-on binary system with a massive evolved sgB[e] star
at the center of the system. This system was shown to be spectroscopically
resolved (in the near infrared) into three distinct emission regions, viz.
\begin{enumerate}
  \item The \refmark{companion star at $\sim$ 20,000~K \citep{raho2008}}
  \item A hot and optically thick rim of puffed up hot dust at $\sim$ 5000~K
	at a distance of $\sim$ 0.8 - 1.2 au. from the companion.
  \item A warm dust viscous disk like shell with inner disk at $\sim$ 800~K
	and extending upto $\sim$ 5.6 au. from the companion.
\end{enumerate}
Given an orbital period of $\sim$ 80.09 days, we can find where the orbit of
the compact object would lie with respect to the companion.
Taking the companion \refmark{mass in the range 25 - 50 M$_\odot$ ( around
the likely companion mass of 30 M$_{\odot}$ reported in \citet{chat2012})} and the compact
object mass in the range of 1.4 - 10 M$_\odot$ (the typical range of masses
of compact objects in our galaxy), we can compute the distance using the
expression for mass function of a binary system. 
The range of mass values assumed, gives a range 
of the separation between the two stars to be in between 0.87 au. to 1.32 au.
\refmark{For a neutron star compact object (mass in the range 1.4 M$_\odot$ to 2.1 M$_\odot$),
the separation comes to be in between 1.0 au to 1.32 au. The small change in
separation between the black hole and neutron star case is because the
separation in a binary system is dependent on the sum of both masses in the
binary. In the case of an HMXB system, this separation is mainly governed by the mass of
the companion, which is much higher than the compact object mass.} 
This geometry is illustrated in Figure \ref{fig:orbit}.

Thus, we find that the compact object spends most of its time in either the hot and
optically thick dust rim or the warm dust disk of the companion. There might be a few orbital
phases when the compact object does enter the cavity in between the
companion and the hot dust rim depending on the eccentricity and the actual
separation between the stars. However, for an observer the dust rim and
the viscous disk always come in the line of sight to the compact object.
Such an orbit would be similar to the binary system geometry used to explain
the high $\mathrm{N_H}$ values in CI Cam \citep[see for e.g. Figure 7 of ][and
references therein]{bart2013}. It is interesting to note that CI Cam too has
a very similar X-ray spectrum to IGR J16318-4848 (in the soft X-rays). These
sources could thus be a separate class of X-ray binaries with
the compact object enshrouded in the dust rim surrounding an sgB[e]
companion.

The two peaks corresponding to phases of enhanced emission can possibly be
linked to in-homogenities in the viscous disk and rim structure surrounding
the central sgB[e] star. The large outburst of Sep 2008 possibly caused some
re-arrangement of material in the hot dust rim, which lead to a large gap
(of 10 orbits) without any detected flare. The re-arranged clumps in the
hot dust rim can be the likely cause of the random distribution of flares
in the Main Peak after the Sep 2008 outburst.

An interesting point worth noting is the absence of cyclotron resonant
scattering features (CRSFs) in the X-ray spectrum of this source. This,
coupled with the absence of pulsation detections raises questions on the type
of compact object. The spectral shape ( which is similar to other neutron
star HMXBs ) and the absence of a radio jet \citep{fill2004} points to a neutron star
nature of the compact object. This system could then possibly be a binary
with the neutron star poles not crossing our line of sight. Even if the
poles do cross our line of sight, smearing
due to scattering \refmark{of X-rays from the} surrounding medium may possibly be wiping out signatures of
pulsations \citep{kust2005}.

\section{Summary and conclusions}\label{sec:sum}
Our analysis confirms an 80.09 day periodicity
in the \textit{Swift} BAT lightcurve of IGR J16318-4848. We take this to be the orbital period
and show the presence of two phases of enhanced X-ray emission in the
orbit folded lightcurve of this source. The 80.09 day orbit places the
compact object in this binary system within a rim of hot dusty gas which
surrounds the companion sgB[e] star. This enshrouded path of the compact
object orbit causes the large observed absorption column in this source. 
Intermittent flares seen in the source could possibly be due to
inhomogeneous sections of the orbit in the hot dust rim. These flares form the 
Main Peak as seen in the orbit folded lightcurve.

Further examination of the spectral variability as a function of the orbital
phase can give a better understanding of this picture. We believe that an orbit
phase resolved observational study of the X-ray spectral parameters
\refmark{needs to be carried out to} improve our knowledge of this interesting binary system.

\section*{Acknowledgements}
This research has made use of data and software provided by the High
Energy Astrophysics Science Archive Research Center (HEASARC), which is a
service of the Astrophysics Science Division at NASA/GSFC and the High
Energy Astrophysics Division of the Smithsonian Astrophysical Observatory.
It is additionally based on observations obtained with XMM-Newton, an ESA science
mission with instruments and contributions directly funded by ESA Member
States and NASA. The research also made use of the Swift/BAT transient monitor results
provided by the Swift/BAT team. We would like to thank all the proposers of the archival
observations used in this work. \refmark{We would also like to thank the
anonymous referee for his/her suggestions, which helped improve the manuscript.}

%%%%%%%%%%%%%%%%%%%%%%%%%%%%%%%%%%%%%%%%%%%%%%%%%%

%%%%%%%%%%%%%%%%%%%% REFERENCES %%%%%%%%%%%%%%%%%%

% The best way to enter references is to use BibTeX:

\bibliographystyle{mnras}
\bibliography{refs} % if your bibtex file is called example.bib

%%%%%%%%%%%%%%%%%%%%%%%%%%%%%%%%%%%%%%%%%%%%%%%%%%

%%%%%%%%%%%%%%%%% APPENDICES %%%%%%%%%%%%%%%%%%%%%

%%%%%%%%%%%%%%%%%%%%%%%%%%%%%%%%%%%%%%%%%%%%%%%%%%

% Don't change these lines
\bsp	% typesetting comment
\label{lastpage}
\end{document}